\documentclass[aps,prl,reprint,groupedaddress,showpacs]{revtex4-1}
\usepackage{amsmath,amssymb}
\usepackage{graphicx}

\begin{document}

\title{Two-Dimensional Superconducting State of Monolayer Pb Films Grown\\
on GaAs(110) in a Strong Parallel Magnetic Field}

\author{Takayuki Sekihara, Ryuichi Masutomi, and Tohru Okamoto}
\affiliation{Department of Physics, University of Tokyo, 7-3-1 Hongo, Bunkyo-ku, Tokyo 113-0033, Japan}

\date{2 August 2013}

\begin{abstract}
Two-dimensional (2D) superconductivity was studied by magnetotransport measurements on
single-atomic-layer Pb films on a cleaved GaAs(110) surface.
The superconducting transition temperature shows only a weak dependence on the parallel magnetic field
up to 14~T, which is higher than the Pauli paramagnetic limit.
Furthermore, the perpendicular-magnetic-field dependence of the sheet resistance is
almost independent of the presence of the parallel field component.
These results are explained in terms of an inhomogeneous superconducting state
predicted for 2D metals with a large Rashba spin splitting.
\end{abstract}

%74.25.F- Transport properties  
%74.78.-w Superconducting films and low-dimensional structures 
%73.20.At Surface states, band structure, electron density of states  
%73.25.+i Surface conductivity and carrier phenomena  
 
%74.20.Mn Nonconventional mechanisms  
%75.70.Ak Magnetic properties of monolayers and thin films  
%75.70.Rf Surface magnetism 
%74.25.Ha Magnetic properties including vortex structures and related phenomena
% (for vortices, magnetic bubbles, and magnetic domain structure, see 75.70.Kw) 

\maketitle

Superconductivity in ultrathin films has been studied for a long time.
In Ref.~\cite{Haviland1989}, superconductivity was observed even for a few-monolayer thickness
in quench-condensed films of Bi and Pb deposited on a glazed alumina substrate coated with amorphous Ge.
Very recently, it has been revealed that superconductivity can occur in single atomic layers of Pb and In
grown epitaxially on a Si(111) substrate \cite{Zhang2010,Uchihashi2011}.
A single-atomic-layer metal film on an insulating substrate is
an interesting system for studies of superconductivity,
not only because it is a complete two-dimensional (2D) system
but also because of the broken spatial inversion symmetry.
The asymmetry of the confining potential in the direction perpendicular to the 2D plane,
combined with atomic spin-orbit coupling,
is expected to cause the Rashba effect, which lifts the spin degeneracy of the 2D electronic states
\cite{Rashba1960,Bychkov1984}.
Actually, angle-resolved photoelectron spectroscopy measurements showed
a large Rashba spin splitting of the order of 100~meV
on the surfaces of heavy elements, such as Au \cite{LaShell1996}, W \cite{Hochstrasser2002},
and Bi \cite{Koroteev2004},
and those of lighter elements, such as Si and Ge, covered with a monolayer of heavy elements,
such as Bi \cite{Gierz2009,Hatta2009,Sakamoto2009} and Pb \cite{Yaji2010}.

In this Letter, we report magnetotransport measurements on
superconducting monolayer Pb films produced by quench condensation onto a cleaved GaAs(110) surface.
Here, we focus on the effect of the magnetic field applied parallel to the surface.
While the perpendicular component $H_\perp$ of the magnetic field strongly affects the orbital motion of electrons in the 2D plane,
the parallel component $H_\parallel$ is expected to couple only to the spin degree of freedom.
We show that the reduction of the superconducting transition temperature $T_c$
is very small even in strong parallel magnetic fields ($H_\perp=0$),
which are much larger than the Pauli paramagnetic limit.
Furthermore, the $H_\perp$ dependence of the sheet resistance at low temperature is found to be
almost independent of the presence of $H_\parallel$.
These results are explained by assuming an inhomogeneous superconducting state
predicted for Rashba spin-split 2D systems.

In order to measure the sheet resistance $R_{sq}$ of ultrathin films,
we apply the experimental procedure developed for studies on adsorbate-induced surface inversion layers
on InAs \cite{Tsuji2005,Mochizuki2008,Okamoto2011} and InSb \cite{Masutomi2007}.
In this work, we used a nondoped insulating GaAs single-crystal substrate
so as not to create conduction channels in the substrate.
Current and voltage electrodes were prepared at room temperature
by deposition of gold films onto noncleaved surfaces.
Cleavage of GaAs, subsequent deposition of Pb, and resistance measurements 
were performed at low temperatures in an ultrahigh vacuum chamber immersed in liquid He.
The deposition amount was measured with a quartz crystal microbalance
and determined with an accuracy of about 5\%.
The four probe resistance of the Pb film  on a cleaved GaAs(110)  surface (4 mm $\times$ 0.35 mm)
was measured using the standard lock-in technique at 13.1 Hz.
The magnetic-field direction, with respect to the surface normal,
was precisely controlled using a rotatory stage on which the sample was mounted,
together with a Hall generator, a RuO$_2$ resistance thermometer, and a heater.
The sample stage can be cooled down to 0.5~K via a silver foil linked to a pumped ${}^3$He refrigerator.
All the data were taken when the temperature of the sample stage was kept constant at a fixed value,
so as to ensure a thermal equilibrium between the sample and the thermometer.
To measure a small change in $T_c$ induced by the magnetic field,
the effect of the magnetic field on the thermometer should be taken into account.
We systematically calibrated the magnetoresistance effect of the RuO$_2$ resistance thermometer
against the vapor pressure of the liquid ${}^3$He or ${}^4$He for various temperatures.
After the correction, $T_c$ can be determined with a relative accuracy of less than 0.2 \% \cite{RuO2}.

Figure 1 shows the zero-field superconducting transition temperature $T_{c0}$
and the normal-state sheet resistance $R_N$
as a function of the nominal thickness of the Pb film.
Data for Bi films are also shown for reference.
We defined the transition temperature as the temperature at which $R_{sq}$ reaches $R_N/2$.
As the film thickness decreases, $R_N$ increases and $T_{c0}$ decreases monotonically
in a similar manner to the data of Ref.~\cite{Haviland1989}.
However, $R_N$ is lower and $T_{c0}$ is higher in this work.
This also holds for Bi films.
We attribute it to an atomically flat surface of the cleaved GaAs substrate.
The lowest nominal thickness of the Pb film for which superconductivity was observed in this work is 0.22~nm,
which corresponds to an atomic areal density of $7.2~\mathrm{nm}^{-2}$.
This value is lower than the surface atom density of GaAs(110)
($8.9~\mathrm{nm}^{-2}$),
the planar density of the bulk Pb(111) plane ($9.4~\mathrm{nm}^{-2}$),
and the atomic areal density in the striped incommensurate phase
of the one-atomic-layer Pb film on Si(111)
($10.4~\mathrm{nm}^{-2}$) \cite{Zhang2010}.
It is unlikely that the second layer promotion occurs at this coverage.

\begin{figure}[t]
\begin{center}
\includegraphics[width=0.92\linewidth, clip]{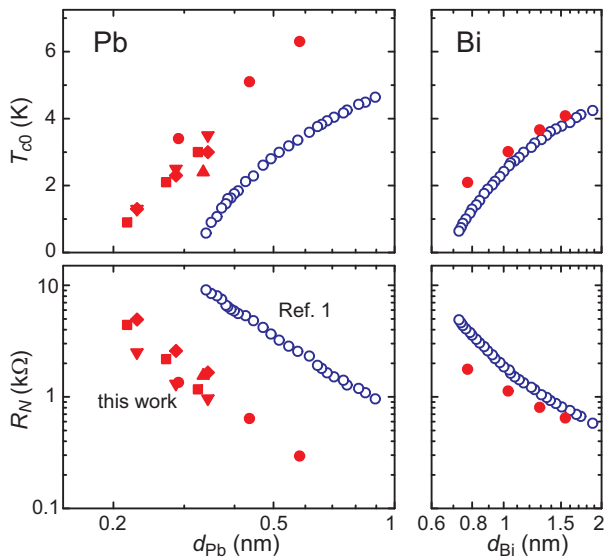}
\caption{(color online) 
Zero-field superconducting transition temperature
and normal-state sheet resistance as a function of
the nominal thicknesses of Pb and Bi films.
Different filled (red) symbols correspond to different experiments in this work.
Open (blue) circles are data from Ref.~\cite{Haviland1989}.
}
\end{center}
\end{figure}

Figure 2 shows the temperature dependence of $R_{sq}$ of Pb films for different atomic areal densities $n$
with and without parallel magnetic field.
Following a gradual decrease with decreasing temperature, $R_{sq}$ falls to zero for all cases.
While $T_c$ depends strongly on $n$, it is almost independent of $H_\parallel$.
In conventional superconductors, the Cooper pair consists of electrons with opposite spins.
It is expected that superconductivity will be destroyed in a strong magnetic field
where the condensation energy is compensated
by the free-energy reduction in the normal state due to the Pauli paramagnetic susceptibility
~\cite{Clogston1962,Chandraskhar1962}.
This limit is known as the Pauli paramagnetic limit.
According to Ref.~\cite{Clogston1962},
the limiting field at $T=0$ is given by $H_P=\Delta_0/\sqrt{2} \mu_B$,
where $\Delta_0$ is the energy gap at $T=0$ and $\mu_B$ is the Bohr magneton.
For a weak-coupling BCS superconductor, it can be expressed as 
$H_P ({\mathrm T}) =1.86 T_{c0} ({\mathrm K})$ \cite{Sacepe2008}.
In the case of Fig.~2, $H_P$ is determined to be	
1.7, 4.8, and 6.1~T for $T_{c0}=0.9$,  2.6, and 3.3~K, respectively.
However, the superconducting state is found to be stable for $H_\parallel>H_P$.
The experimental results indicate that the critical field $H_c$, if it exists, is much higher than $H_P$.

\begin{figure}[t]
\begin{center}
\includegraphics[width=0.85\linewidth, clip]{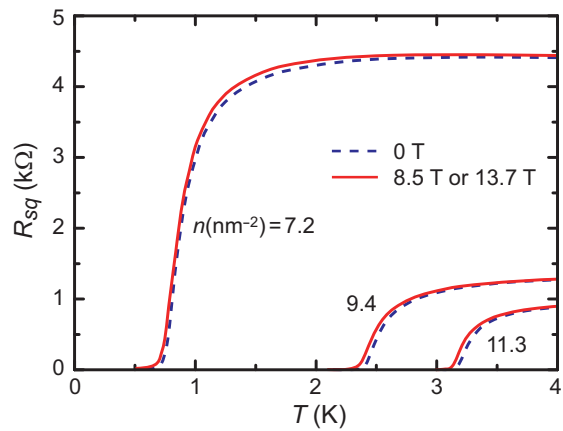}
\caption{(color online)
$T$ dependence of the sheet resistance of Pb films for different atomic areal densities.
Dashed (blue) curves are obtained at zero magnetic field.
Solid (red) curves are obtained in parallel magnetic field of
8.5~T (for $n=7.2~\mathrm{nm}^{-2}$)
or 13.7~T (for $n=9.4$ and $11.3~\mathrm{nm}^{-2}$).
The magnetic-field angle was adjusted to $H_\perp=0$
using the $R_{sq}$ minimum.
}
\end{center}
\end{figure}

In contrast to the case of $H_\parallel$,
the superconducting state is easily destroyed by $H_\perp$ because of the orbital effect.
The solid curve in Fig.~3 represents the perpendicular-magnetic-field dependence
of $R_{sq}$ of the Pb film at $n=7.2~\mathrm{nm}^{-2}$ and $T=0.51$~K.
\begin{figure}[b]
\begin{center}
\includegraphics[width=0.85\linewidth, clip]{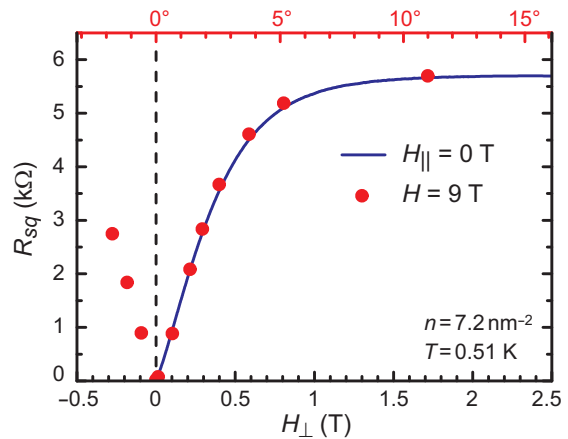}
\caption{(color online)
$H_\perp$ dependence of the sheet resistance of the Pb film at $n=7.2~\mathrm{nm}^{-2}$ and $T=0.51$~K.
The solid (blue) curve is obtained for the perpendicular-magnetic-field direction ($H_\parallel=0$).
Filled (red) circles are obtained by changing the magnetic-field angle (upper axis)
at a fixed strength of 9~T.
}
\end{center}
\end{figure}
In low fields, $R_{sq}$ increases almost linearly with $H_\perp$,
indicating that vortex pinning, which arises from structural inhomogeneities,
is very weak in the Pb film.
From the $T$ dependence of $R_{sq}$ (not shown here),
the critical field for the magnetic-field-induced superconductor-insulator transition
was found to be 0.6~T.
If we use this value as the upper critical field,
the Ginzburg-Landau coherence length is calculated to be 23~nm.
In order to study the effect of $H_\parallel$,
another experiment was performed at a fixed magnetic-field strength of 9~T.
The perpendicular component $H_\perp$ was introduced by 
changing the magnetic-field angle from the parallel direction.
The results, plotted as filled circles in Fig.~3,
show good agreement with those for $H_\parallel=0$.
This also demonstrates that a parallel magnetic field
exceeding several times the Pauli paramagnetic limiting field
does not have a strong effect on the stability of the superconducting state.

It is well known that the effect of the spin-orbit scattering
can lead to an enhancement of the critical field.
According to the theory for a superconductor with strong spin-orbit scattering \cite{Maki1966,Klemm1975}, 
$H_c$ is given by
\begin{equation}
\ln \left( \frac{T}{T_{c0}} \right)+\psi \left( \frac{1}{2} +
\frac{3 \tau_\mathrm{SO} (\mu_B H_{c})^2 }{4 \pi \hbar k_B T}
\right) - \psi \left( \frac{1}{2}\right) =0,
\end{equation}
where $\tau_\mathrm{SO}$ is spin-orbit scattering time and $\psi(x)$ is the digamma function.
In the zero-temperature limit, we obtain
\begin{equation}
H_{c}
=\frac{1}{\mu_B} \sqrt{ \frac{\pi \hbar k_B T_{c0}}{ 3  e^\gamma \tau_\mathrm{SO}} }
=0.602 \sqrt{\frac{\hbar}{ \tau_\mathrm{SO} k_B T_{c0}} }H_P,
\end{equation}
where $\gamma=0.5772\ldots$ is Euler's constant \cite{EulersConstant}.
For $\hbar \tau_\mathrm{SO}^{-1} \gg k_B T_{c0}$,
$H_c$ can be much higher than $H_P$.
On the other hand, for small magnetic fields $H_\parallel \ll H_c (T=0)$,
the $H_\parallel$ dependence of $T_c$ is obtained from Eq.~(1) as
\begin{equation}
T_c=T_{c0}- \frac{3 \pi \tau _{so} \mu _B^2}{8 \hbar k_B} H_\parallel^2.
\end{equation} 
In Fig.~4, experimental results for $\Delta T_c \equiv T_c-T_{c0}$ are shown.
The superconducting transition temperature exhibits a parabolic $H_\parallel$ dependence,
which is at least qualitatively consistent with Eq.~(3).
If we apply Eq.~(3) to the experimental data,
$\tau_\mathrm{SO}$ is determined to be 3.5 and 4.3~fs
for $n=9.4$ and $11.3~\mathrm{nm}^{-2}$, respectively.
The dominant contribution to electron scattering is expected to
come from defects in our quench-condensed Pb films.
Since electrons cannot move in the direction perpendicular to the surface in a monolayer film,
we do not consider here surface scattering
which can be important in moderately thin films and fine particles.
In Ref.~\cite{Belevtsev1998},
the probability of the spin-orbit process on elastic (momentum) scattering 
was determined for defect scattering in 10-nm-thick Au films after irradiation of Ar ions.
From weak localization magnetoresistance measurements,
the ratio of $\tau_\mathrm{SO}^{-1}$ to the elastic scattering rate $\tau^{-1}$
was found to be $4.0 \times 10^{-4}$.
For the present Pb films,
$\tau$ is roughly estimated from $R_N$ to be 3.1 and 3.5~fs
for $n=9.4$ and $11.3~\mathrm{nm}^{-2}$, respectively \cite{ScatteringTime}.
If the above values of $\tau_\mathrm{SO}$ are substituted,
we obtain $\tau/\tau_\mathrm{SO} \sim 0.8$, which is 3 orders of magnitude larger than that in the Au films.
While the strength of the spin-orbit interaction strongly depends on the atomic number $Z$,
the difference between Au ($Z=79$) and Pb ($Z=82$) is small.
It seems unreasonable to expect such a large value of $\tau/\tau_\mathrm{SO}$ for the Pb films.

\begin{figure}[tb]
\begin{center}
\includegraphics[width=0.85\linewidth, clip]{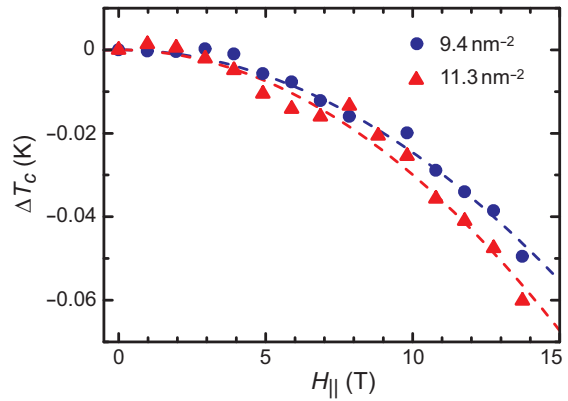}
\caption{(color online) 
$H_\parallel$ dependence of $\Delta T_c \equiv T_c-T_{c0}$
for Pb films with $n=9.4$ and $11.3~\mathrm{nm}^{-2}$.
The dashed lines are the best parabolic fits.
}
\end{center}
\end{figure}

We now consider the 2D superconducting state
in the presence of a Rashba spin splitting.
Because of the broken inversion symmetry
and the strong atomic spin-orbit coupling of Pb,
a large splitting is expected in our Pb films
as in other monolayer systems \cite{Gierz2009,Hatta2009,Sakamoto2009,Yaji2010}.
The Rashba interaction couples the electron spin to the momentum.
The Cooper pairs are not in a pure spin-singlet state
but in a mixture of singlet and triplet states
\cite{Edelstein1989,Gorkov2001}. 
It was shown in~Ref.~\cite{Barzykin2002} that
the critical field in the parallel direction can be very high
because of a formation of an inhomogeneous superconducting state 
similar to the Fulde-Ferrel-Larkin-Ovchinnikov state
\cite{Fulde1964,Larkin1965}.
In this state, which is also called a stripe state, the two Fermi circles are shifted
in opposite $y$ directions by a momentum $q/2=\mu_B H_\parallel/v_F$ in a magnetic field along the $x$ direction,
and the order parameter varies as $\cos (qy)$.
Here, $v_F$ is the Fermi velocity.
The critical field is given by
\begin{equation}
H_{c}
=\frac{1}{\mu_B} \sqrt{ \Delta_0 \Delta_R}
=\frac{1}{\mu_B} \sqrt{ \frac{\pi k_B T_{c0} \Delta_R}{e^\gamma}},
\end{equation}
where $\Delta_R$ is the Rashba spin splitting $\Delta_R \gg k_B T_{c0}$.
For $\Delta_R=0.1$~eV and $T_{c0} =3$~K, we obtain $H_c \sim 100$~T.

It has been demonstrated that the Fulde-Ferrel-Larkin-Ovchinnikov-like phase is eliminated by
a sufficiently high concentration of nonmagnetic impurities
~\cite{Dimitrova2003,Dimitrova2007}.
Instead, there appears a long-wavelength helical state
where the Cooper pairs have a nonzero momentum $q = \mu_B H_\parallel \Delta_R/v_F \epsilon_F$
and the order parameter varies as $\exp (i qy)$.
Here, $\epsilon_F$ is the Fermi energy.
For $\Delta_R \gg \hbar \tau^{-1}$, the paramagnetic critical field is given by
\begin{equation}
H_{c}=\frac{1}{\mu_B} \sqrt{ \frac{\pi \hbar k_B T_{c0}}{4e^\gamma \tau}}.
\end{equation}
Note that, unlike Eq.~(2), the elastic scattering time determines $H_c$.
The Rashba interaction allows nonmagnetic scattering to mix spins.
We expect the theory to be applicable to our Pb films
where defect scattering is likely to be the dominant scattering mechanism.
For small $H_\parallel$, the $H_\parallel$ dependence of $T_c$ can be expressed as
\begin{equation}
T_c=T_{c0}- \frac{\pi \tau \mu _B^2}{2 \hbar k_B} H_\parallel^2.
\end{equation} 
From the data in Fig.~4,
we obtain $\tau = 2.6$ and 3.2~fs
for $n=9.4$ and $11.3~\mathrm{nm}^{-2}$, respectively.
These values are in good agreement with those estimated from $R_N$.
Thus, the theory successfully accounts for the robustness of superconductivity against $H_\parallel$
and the parabolic $H_\parallel$ dependence of $T_c$.

The effects of Rashba spin splitting on superconducting-state properties have been discussed
for the SrTiO${}_3$/LaAlO${}_3$ interface \cite{Shalom2010}
and Nb-doped SrTiO${}_3$ heterostructures \cite{Kim2012}.
For both systems, $H_c$ exceeding $H_P$ by a factor of 4 was reported.
However, the theories for 2D Rashba superconductors
\cite{Edelstein1989,Gorkov2001,Barzykin2002,Dimitrova2003,Dimitrova2007}
cannot be applied to these cases
because electrons are distributed in the conduction layer of several nm thickness
and many subbands are expected to be occupied.

In summary, we have studied 2D superconductivity in ultrathin Pb films
on a cleaved GaAs(110) surface.
Zero resistance was clearly observed even for 0.22~nm nominal thickness,
which corresponds to (sub)monolayer coverage.
The reduction of $T_c$ was found to be very small
even in parallel magnetic fields exceeding several times the Pauli paramagnetic limit.
Furthermore, tilted-magnetic-field measurements show that
the $H_\perp$ dependence of the sheet resistance at low temperature is
almost independent of the presence of $H_\parallel$.
The spin-orbit scattering can account for the observed robustness of superconductivity against $H_\parallel$
only if a large value of $\tau/\tau_\mathrm{SO}$ is assumed.
On the other hand, the results are consistently explained
in terms of an inhomogeneous superconducting state predicted for 2D metals with a large Rashba spin splitting.

\begin{acknowledgments}
We are grateful to S. Fujimoto and T. Kanao for helpful suggestions.
This work has been partially supported 
by a Grain-in-Aid for Scientific Research(A)(No.21244047) from MEXT, Japan.
\end{acknowledgments}


\begin{thebibliography}{99}
\bibitem{RuO2}
The RuO$_2$ resistance thermometer (RO-600) is manufactured by Scientific Instruments, Inc.,
which offers a simple linear formula for the positive magnetoresistance.
It allows correction of any apparent temperature,
measured in a magnetic field up to 16~T, to the actual temperature
with a temperature error of less than $\pm 1.6$\%
for a broad temperature range of 0.036 to 4.2~K.
Our calibration agrees with the formula within the error bars.
\bibitem{EulersConstant}
Note that $\gamma$ denotes $e^\gamma=1.78$ in some references.
\end{thebibliography}
\end{document}